\documentclass[prl,twocolumn,showpacs]{revtex4-1}

\usepackage{color}
\usepackage{amsfonts}
\usepackage{amssymb}
\usepackage{amsthm}
\usepackage{hyperref}
\usepackage{graphicx}

\def\be{\begin{equation}}
\def\ee{\end{equation}}
\def\bea{\begin{eqnarray}}
\def\eea{\end{eqnarray}}
\definecolor{comment}{rgb}{0.5,0.5,0}
\begin{document}

\title{Why is lead so kinky?} 
\author{P. M. Goddard, P. D. Stevenson and A. Rios}
\affiliation{Department of Physics, University of Surrey, Guildford GU2 7XH, United Kingdom}

\begin{abstract}
We revisit the problem of the kink in the charge radius shift of neutron-rich even isotopes near the $N=126$ shell closure. We show that the ability of a Skyrme force to reproduce the isotope shift is determined by the occupation of the neutron $1i_{11/2}$ orbital beyond $N=126$ and the corresponding change it causes to deeply-bound protons orbitals with a principal quantum number of $1$. 
Given the observed position of the single-particle energies, one must either ensure occupation is allowed through correlations, or not demand that the single-particle energies agree with experimental values at the mean-field level.
\end{abstract}
	
\pacs{21.10.Ft, 21.60.Jz, 21.30.-x, 27.80.+w}

\maketitle

The evolution of charge radii across the isotope chart is one of the most basic nuclear structure observables, as it 
provides a particularly useful characterisation of the proton distribution that can be accessed by a variety of experiments \cite{Angeli}.  
We define the charge radius isotope shift as the differences between the mean squared charge radius, $\langle r^2_{\textrm{ch}} \rangle$, of a series of isotopes and that of a given reference isotope ($^{208}$Pb for lead and $^{210}$Po for polonium).
Theoretically, charge radii have been traditionally studied within the droplet model \cite{Myers1983}, which captures qualitatively their evolution across the nuclear chart. Yet, in some specific cases, quantum shell effects dominate the density distribution and provide a departure from smooth systematic trends. Perhaps the most well-known example of these abrupt changes is the kink in the isotope shift of even lead isotopes as one passes through the $N=126$ magic number.

\begin{figure}[tb]
\includegraphics[width=\linewidth]{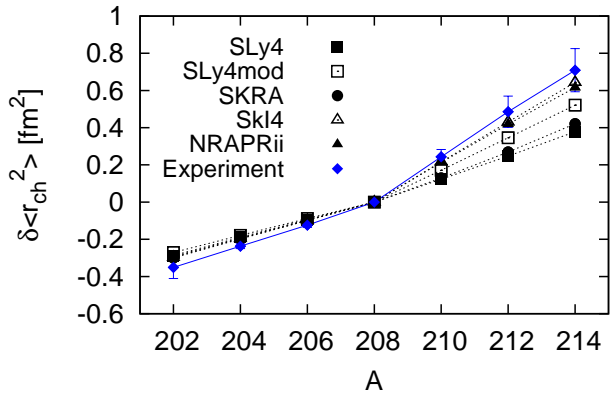}
\caption{Isotope shifts are given by the difference in the mean square charge radius between a series of even isotopes, denoted by their mass number $A$, and that of $^{208}$Pb. Across the $N=126$ shell closure, a strong increase in the slope of the experimental data (diamonds) is observed. Theoretical predictions, obtained with different Skyrme parametrizations, are also presented. Only a handful of these Skyrme sets are able to reproduce the increase of slope above $N=126$.}
\label{fig:pbshift}
\end{figure}

\begin{figure}[tb]
\includegraphics[width=\linewidth]{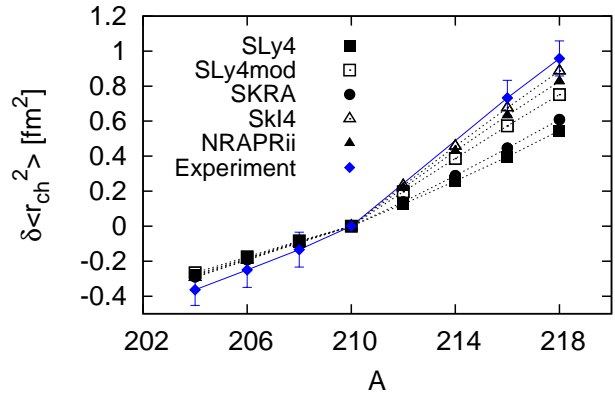}
\caption{The same as Fig.~\ref{fig:pbshift} for polonium. The reference isotope is $^{210}$Po.}
\label{fig:poshift}
\end{figure}

A summary of the experimental results \cite{Angeli,Cocolios2011} around this neutron shell closure is given in Figures \ref{fig:pbshift} and \ref{fig:poshift} for lead and polonium, respectively. For lead, the fitted solid line is of the form:
\be
\delta \langle r^2_{\textrm{ch}} \rangle 
= \left\{\begin{array}{lr}m_1(A-208),&A<208\\m_2(A-208),&A>208\end{array}\right.,
\ee
Linear regression gives
\bea
m_1 &=& 0.0598(7)\ \mathrm{fm^2}\nonumber \\
m_2 &=& 0.1203(9)\ \mathrm{fm^2}\nonumber,
\eea
i.e. the slope of the shift is observed to double at $A=208$ to a very good approximation. 
This abrupt change in change radius cannot be explained within the droplet model \cite{Myers1983}. 
Recent experimental work using laser spectroscopy techniques have identified a similar kink in  neutron-rich polonium isotopes above the $Z=82$ shell closure \cite{Cocolios2011} (see also Fig.~\ref{fig:poshift}) in agreement with older radon and radium data \cite{Otten}.
Moreover, the details of the changes of proton and neutron radii in isotopes around $^{208}$Pb have also gained renewed interested in the form of the neutron skin, which correlates strongly with nuclear matter properties \cite{roca2011,prex}.
On the neutron deficient side, the onset of deformation is also probed by measurements of isotope shifts \cite{DeWitte2007}. 
In this letter, we propose a new mechanism to explain the existence of the kink in lead isotope shifts using density functional calculations supplemented by pairing effects.

On the theoretical side, mean-field models, or equivalently density functional theories have been widely applied to the systematic study of all observed and hypothesized nuclear isotopes \cite{Erler2012}.  
Both the Skyrme-Hartree-Fock (SHF) and the Relativistic Mean-Field (RMF) approaches are able to give broadly good descriptions of many nuclear properties across the nuclear chart, including radii \cite{BenderReview}. In particular, shell effects are naturally included in such quantum mechanical calculations. The kink in the isotope shift for lead, however, is still a somewhat challenging observable, for which a full theoretical understanding remains elusive \cite{Cocolios2011}. 
Note, in particular, that beyond mean-field correlations have a very small effect on the charge radii in this area, close to the shell closure \cite{Bender2006}. This suggests that any missing theoretical component should have a rather general origin, most likely independent of the particular microscopic picture.

The ability of different mean-field methods to reproduce (or not) the isotope shift in lead has often been used to benchmark different models \cite{Tajima,Reinhard,StoneReview}. Original Skyrme parameterizations were unable to reproduce the isotope shift in lead, whereas RMF parameterizations seemed to be able to do so.  Since the early 1990s, the situation has been studied by several authors \cite{Tajima,Reinhard,Bender2006,Niembro2012}. Sharma {\it et al.} \cite{Sharma} and Reinhard and Flocard \cite{Reinhard} took the hint that the spin-orbit mean-field resulting from the SHF and RMF approaches has a different isotopic dependence \cite{Ring2012}.  Reinhard and Flocard extended the Skyrme approach with an extra degree of freedom in the spin-orbit channel \cite{Reinhard}.  In their notation, the Energy Density Functional (EDF) due to the spin-orbit interaction can be written:
\begin{equation}
\epsilon_{ls} = \int\,d^3r\,\bigg\{b_4\rho\nabla\mathcal J + \sum_{q\in\left\{p,n\right\}}b'_4\rho_q\nabla\mathcal J_q\bigg\}. 
\label{eq:soforce}
\end{equation}
with $\rho$ denoting the total particle density, $\rho_q$, the proton (or neutron) density, $\mathcal J$, the spin-orbit current for both types of particle and $\mathcal J_q$, that for a particular type.

With the choice $b_4=b'_4$, one retrieves the form from the original Skyrme force as posited by Bell and Skyrme \cite{BellSkyrme,Skyrme}, while for $b'_4=0$ one has an RMF-type spin-orbit mean field.  One need not apply either of these rules, but rather allow a free fit of both $b_4$ and $b'_4$.  Indeed, the best $\chi^2$ found in Ref.~\cite{Reinhard} -- the widely-applied force SkI4 -- had $b_4\simeq -b'_4$.  Furthermore, it was found that a qualitative description of the kink could be found with the original Bell-Skyrme spin-orbit functional if one included the isotopic shift data in the fit. This, however, can only be achieved at the expense of the overall fit quality through an unusually low effective mass, as is the case for force SkI5 \cite{Reinhard}. 

The usual explanation given for the reproduction (or lack of) of the kink is the position of the $2g_{9/2}$ neutron orbital above the N=126 shell gap \cite{Reinhard,Klupfel}, which in turn is particularly sensitive to the spin-orbit component of the EDF. If this orbital is subject to a weak spin-orbit force, it will not be so deeply-bound and its radius will be commensurately larger. Consequently, the pull on the proton states through the symmetry energy will be somewhat pronounced and might be able to explain the sudden change in charge radius. 
It has also been noted that the $1i_{11/2}$ neutron state is sometimes occupied through pairing (typically BCS) probabilities. Yet, the $1i_{11/2}$ orbital is consistently $\sim$1 fm smaller in radius than the $2g_{9/2}$ \cite{Reinhard}, so a description based on the radius of occupied neutron orbitals cannot give a full explanation. In this letter, we seek to clarify the important role of the occupation of the neutron $1i_{11/2}$ orbital in producing the correct isotope shift. We advocate that the overlap between neutron and proton orbitals with the same principal quantum number, $n=1$, plays a major role. 

The motivation for this study comes partly from a recent evaluation of most known Skyrme parameterizations for their ability to fit nuclear matter properties \cite{Dutra}. 
We would like to see how those forces that passed a series of nuclear matter constraints perform in finite nuclei \cite{brazil}. Note that the spin-orbit term of the mean field is inactive in infinite nuclear matter. Consequently, functionals which perform well in infinite systems might not be able to reproduce observables like the isotope shift. 
A slight subset of the ``good'' forces is used in the present work to serve as an example. 
They are:  
SKRA \cite{skra}, fitted to a realistic nuclear matter equation of state (EoS) along with some finite nuclear properties; and 
NRAPR \cite{NRAPR}, fitted to the EoS, with adjustment of the spin-orbit force to optimise binding energy and radii in some doubly-magic nuclei.
We also consider the well known forces
SLy4 \cite{Chabanat1998}, a widely-used parametrization from the Lyon group; 
and the already mentioned SkI4 \cite{Reinhard}, explicitly adjusted to reproduce isotope shifts in lead.
We note that only one of these parameter sets, SkI4, uses the extended spin-orbit force of Eq.~(\ref{eq:soforce}). In addition, we re-adjust SLy4, as stated below, to analyse the case of the extended form.  We would also like to stress that our conclusions are general and do not depend on the specific Skyrme functional at hand.

Let us briefly review the spin-orbit mean field properties of some of the Skyrme models used here. As it stands, the spin-orbit parameter for NRAPR, $W_0=2b_4=2b_4'=41.958$ MeV fm$^{5}$ is very low (typical values for other forces are $W_0\simeq 120$ MeV fm$^5$).  This results, in particular, in a proton shell-gap so small at Z=82 as to destroy the doubly-magic nature of the nucleus.  We therefore replace NRAPR by constructing a modified force, NRAPRii, in which the value of the $W_0$ is doubled, leaving other parameters unchanged. 
Further, to analyse the effect of a RMF-type spin-orbit force, we introduce a modified SLy4 force, with the original parameters ($b_4=b'_4=61.5$ MeV fm$^5$) changed to $b_4=75.0$ MeV fm$^5$ and $b'_4=0.001$ MeV fm$^5$. The change in $b_4$ for SLy4mod was designed to keep the magnitude of the spin-orbit field in $^{208}$Pb the same as SLy4, while altering only its isovector properties. This only modifies total energies and radii of closed shell nuclei within $2 \%$. 

The calculated isotope shifts for these five Skyrme parameterizations in lead and polonium are shown in Figs.~\ref{fig:pbshift} and Fig.~\ref{fig:poshift}, respectively. Calculations were performed with a spherical Hartree-Fock code, including BCS pairing with a delta volume pairing force \cite{BenderReview}.
Deformed calculations for $^{186}$Pb, not included here for brevity, indicate that all these forces also provide a plausible description of non-spherical isotopes.
SLy4mod, NRAPRii and SkI4 are able to produce a sizable kink in the isotope shift. Note, however, that SkI4 does so by construction. Most Skyrme forces, including SLy4 and SKRA, struggle to reproduce the kink. 
The forces which have a kink in lead do produce one in polonium. The mechanism underlying the kink should therefore be quite general within this mass region.

\begin{figure}[tb]
\includegraphics[width=\linewidth]{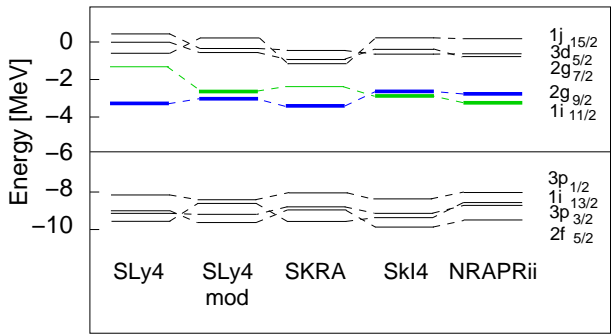}
\caption{Neutron single-particle energies around the Fermi surface in $^{210}$Pb for five sets of Skyrme forces. States with a significant BCS occupation, $>3 \%$, above N=126 are emboldened \label{fig:nspe}. Whenever the $1i_{11/2}$ state is substantially populated, the kink in isotope shift can be reproduced.}
\end{figure}

A key difference between the forces showing the kink and the others is illustrated in Figure \ref{fig:nspe}.  This shows the neutron single-particle levels in $^{210}$Pb, which are only slightly rearranged from those in $^{208}$Pb.  In particular, the bolding of the levels above the N=126 shell closure indicates those orbits outside the $^{208}$Pb core with a BCS occupation above $3 \%$ \footnote{In most cases, the next more occupied level, either the $2g_{7/2}$ or the $1j_{15/2}$ orbitals, has an occupation which is less than $1 \%$.}.
A key point of this Letter is the observation that the $1i_{11/2}$ state is {\it substantially} occupied only in the forces which reproduce the kink. The occupation of this state is vital for getting a kink in other isotopic chains. 

\begin{figure*}[tbh]
\includegraphics[width=\linewidth]{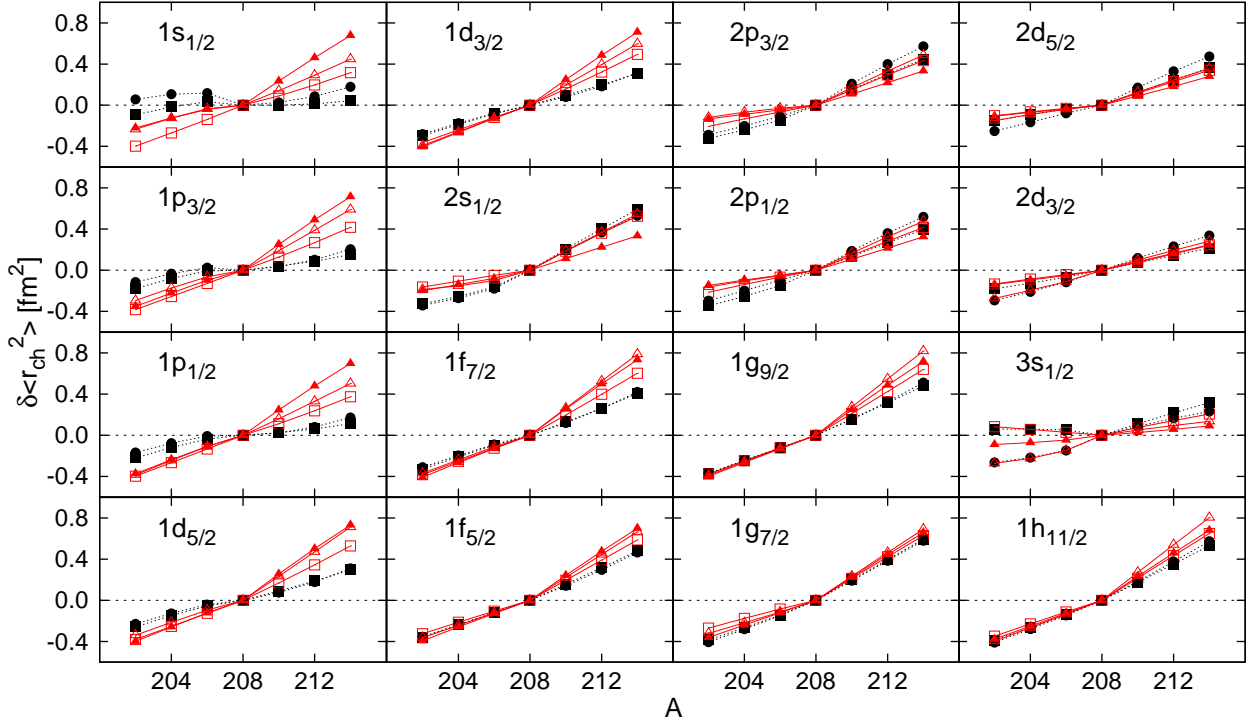}
\caption{Lead isotope shifts of individual proton orbitals  for five sets of Skyrme forces.
Filled squares correspond to SLy4, open squares to SLy4mod, filled circles to SKRA, open triangles to SkI4 and filled triangles to NRAPRii. The results highlighted in red are those associated to forces which reproduce the isotope shift. Above the shell closure, the $n=1$ states in the leftmost column contribute substantially to the kink. 
\label{fig:porbitals}}
\end{figure*}

The effect of the $1i_{11/2}$ occupation on the charge radii is explored in Figure \ref{fig:porbitals}.  Each frame shows the radius isotope shift of every individual occupied proton orbitals from the lowest $1s_{1/2}$ state up to the $1h_{11/2}$ orbital below the $Z=82$ shell gap.  For the forces in which the $1i_{11/2}$ neutron orbital has a significant occupation above $N=126$, there is an increase in the isotope shift of the proton states with principal quantum number $n=1$. States with higher principal quantum numbers show a more mixed behaviour. $n=2$ states have less marked differences across forces, while the parameterization dependence of the $3s_{1/2}$ state is concentrated on the $A<208$ side. The combined contribution of the $n=1$ states explains the presence of the kink in the isotope shift \cite{Niembro2012}.

\begin{figure}[tb]
\includegraphics[width=\linewidth]{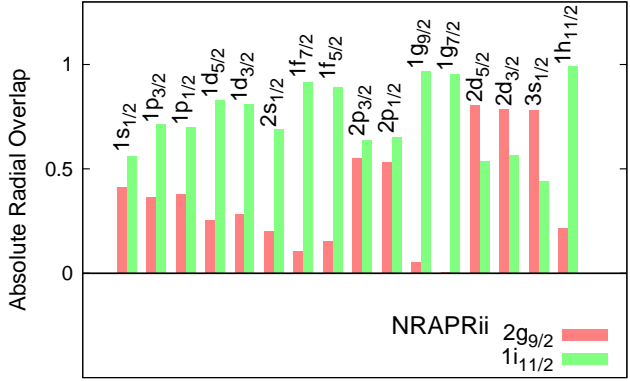}
\caption{Radial overlaps between all the occupied proton orbitals in $^{208}$Pb and the neutron $2g_{9/2}$ (left bars) and $1i_{11/2}$ (right bars) states computed with the NRAPRii parametrization. An overall predominance of overlaps where both states have $n=1$ is found. 
\label{fig:overlaps}}
\end{figure}

From a nuclear matter perspective, one can say that the strong nuclear symmetry energy acts to increase the overlap between all the proton states and the overall nuclear density. The effect will be enhanced if the overlap between wave functions is maximal, which occurs whenever the nodal structure is the same. The latter is determined by the principal quantum number. 
To test this, we calculate the radial overlaps in $^{208}$Pb between the (unoccupied) $1i_{11/2}$ and $2g_{9/2}$ neutron orbitals with each of the occupied proton orbitals.  Since the particles in each orbital interact via the nodeless mean field, and because of the symmetry energy tending to favour overlapping states, the radial overlaps serve as a useful proxy for understanding the isotope shift.  These are shown in Figure \ref{fig:overlaps} for NRAPRii, but similar results hold for the other Skyrme forces. A clear dominance of the $1i_{11/2}$ overlaps over the $2g_{9/2}$ ones is seen.  Hence, when neutrons are added to the $1i_{11/2}$ state, the proton states are attracted to larger radii to maximally overlap with the extra neutrons.

This leaves the question of the  observed locations of the two neutron orbitals in question.  Experimentally, the $2g_{9/2}$ orbital is more deeply bound than the $1i_{11/2}$ orbital \cite{brown}, though separated only by $~1$ MeV. SLy4mod, SkI4 and NRAPRii predict a smaller (or even a negative) splitting between these orbitals. If one identifies the mean-field single-particle energies with experimentally measured values, the mechanism we propose can only work if the pairing interaction allows for the scattering of enough Cooper pairs of neutrons into the $1i_{11/2}$ orbital. In general, however, mean-field single-particles energies should not necessarily agree with measured nuclear energy levels \cite{Duguet2012}.

As previously observed, forces with an RMF-style spin-orbit force, like SkI4, or a low effective mass, also tend to give a level ordering favouring the occupation of the $1i_{11/2}$ neutron orbit. According to our explanation, this should lead to a good reproduction of the lead isotope shifts. Our observation also gives scope for a) forces in which energy levels themselves are fitted, thus leading to effective masses close to $m^*/m=1$ \cite{brown} and providing correlations that allow for a sufficient occupation of the $1i_{11/2}$ orbital, or b) adjustments of the single-particle levels via the tensor force \cite{emma} to reproduce correctly this benchmark, without damaging good nuclear matter properties.

In conclusion, we have demonstrated that the reproduction of the isotope shift in lead is by and large determined by the occupation of the $1i_{11/2}$ neutron orbital. 
Since this is an $n=1$ orbital, it overlaps more strongly with the majority of the proton orbitals, including those that are deeply-bound. This provides a larger ``pull" of neutron states on proton orbits via the symmetry energy and allows for the reproduction of the well-known kink across the $N=126$ shell gap. Our explanation is in contrast to the more traditional account, based on the effect of the $2g_{9/2}$ on the proton density. 
For our calculations, we have tweaked the spin-orbit component of a few mean-field parametrizations to lower the $1i_{11/2}$ single-particle energy and increase its occupation. The physical origin of this occupation, however, is not important for the general mechanism presented here.

\acknowledgments

This work has been supported by STFC grants ST/J00051/1, ST/J500768/1, and ST/I005528/1.

\bibliography{biblio}

\end{document}